\documentclass{svproc}

\usepackage{url,graphicx,amsmath}

\begin{document}
\mainmatter              
\title{Evolution of fluctuations in the initial state \\ of heavy-ion collisions from RHIC to LHC}
\titlerunning{Evolution of initial-state fluctuations from RHIC to LHC}  
%
\author{Giuliano Giacalone\inst{1}, Franc\c cois Gelis\inst{1}, Pablo Guerrero-Rodr\'iguez\inst{2}, \\ Matthew Luzum\inst{3}, Cyrille Marquet\inst{4}, Jean-Yves Ollitrault\inst{1}}

\institute{ Institut de physique th\'eorique, Universit\'e Paris Saclay,  \\ CNRS, CEA, F-91191 Gif-sur-Yvette, France \\
\and
Department of Physics, University of Jyv\"askyl\"a, \\ P.O. Box 35, 40014 University of Jyv\"askyl\"a, Finland
\and  
Instituto de F\'{i}sica, Universidade de S\~{a}o Paulo, 
\\ R. do Mat\~{a}o 1371, 05508-090  S\~{a}o Paulo, SP, Brazil
\and
CPHT, CNRS, \'Ecole Polytechnique, Institut Polytechnique de Paris, \\ Route de Saclay, 91128 Palaiseau, France
}

%
%

\tocauthor{Cyrille Marquet}

\authorrunning{Giuliano Giacalone \textit{et al.}} 

\maketitle              

\begin{abstract}
Fluctuations in the initial state of heavy-ion collisions are larger at RHIC energy than at LHC energy. This fact can be inferred from recent measurements of the fluctuations of the particle multiplicities and of elliptic flow performed at the two different energies. We show that an analytical description of the initial energy-density field and its fluctuations motivated by the color glass condensate (CGC) effective theory predicts and quantitatively captures the measured energy evolution of these observables. The crucial feature is that fluctuations in the CGC scale like the inverse of the saturation scale of the nuclei.

\end{abstract}

Data collected at the CERN Large Hadron Collider (LHC) and at the BNL Relativistic Heavy Ion Collider (RHIC) indicates that initial-state fluctuations in relativistic heavy-ion collisions are larger at RHIC. Two observables support this statement: The relative fluctuations of the charged-particle multiplicity, $N_{\rm ch}$, in central collisions, and the relative fluctuations of elliptic flow, $v_2$.
\begin{figure}[t]
    \centering
    \includegraphics[width=\linewidth]{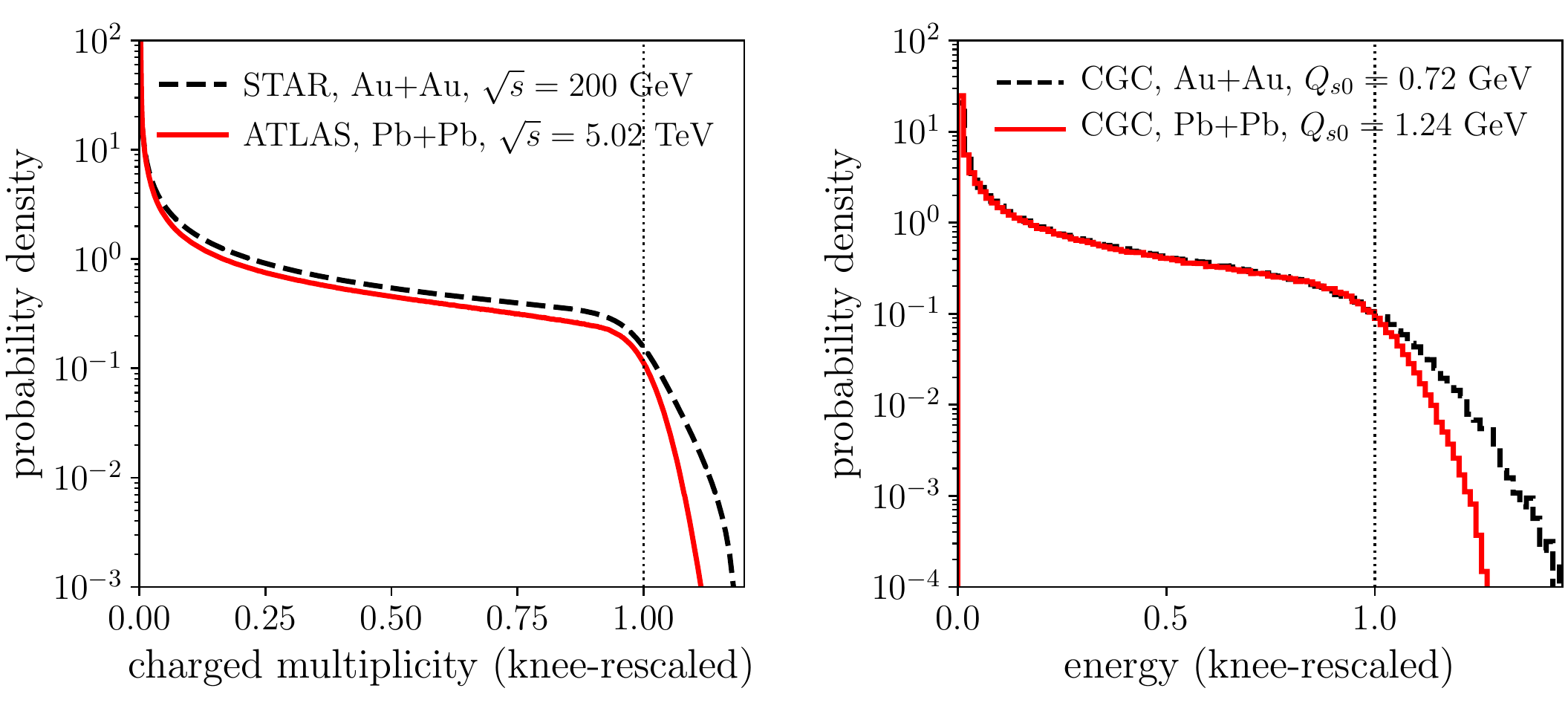}
    \caption{Left: Rescaled histograms of the charged-particle multiplicity measured by the STAR Collaboration~\cite{Adamczyk:2015obl} and by the ATLAS Collaboration~\cite{Aaboud:2019sma}. Right: Rescaled histogram of the total energy in the initial state of Pb+Pb and Au+Au collisions obtained in the CGC model (using the Monte Carlo implementation of Ref.~\cite{Gelis:2019vzt}).}
    \label{fig:1}
\end{figure}

Fluctuations of $N_{\rm ch}$ in nucleus-nucleus collisions probe the fluctuations of the initial state because they are to a good approximation equal to the relative fluctuations of initial total entropy of the system.
The left panel of Fig.~\ref{fig:1} shows the distribution of charged multiplicity measured by the STAR Collaboration in Au+Au collisions at $\sqrt{s}=200$~Gev~\cite{Adamczyk:2015obl}, and by the ATLAS Collaboration in Pb+Pb collisions at $\sqrt{s}=5.02$~TeV~\cite{Aaboud:2019sma}.
The histograms are here rescaled by their value at the \textit{knee}, i.e., the mean value of $N_{\rm ch}$ at zero impact parameter, inferred with the Bayesian procedure of Ref.~\cite{Das:2017ned}. The fluctuations of $N_{\rm ch}$ around the knee quantify the width of the large-multiplicity tail, and this is 1.5 times larger in STAR data:
\begin{equation}
\label{eq:mfluct}
    \frac{\sigma[N_{\rm ch}](b=0)}{\langle N_{\rm ch} \rangle (b=0)} \biggl|_{\rm RHIC} = 0.065, \hspace{50pt}  \frac{\sigma[N_{\rm ch}](b=0)}{\langle N_{\rm ch} \rangle (b=0)} \biggl|_{\rm LHC} = 0.044. 
\end{equation}
This implies that initial-state fluctuations are larger at RHIC.

\begin{figure}[t!]
    \centering
    \includegraphics[width=.55\linewidth]{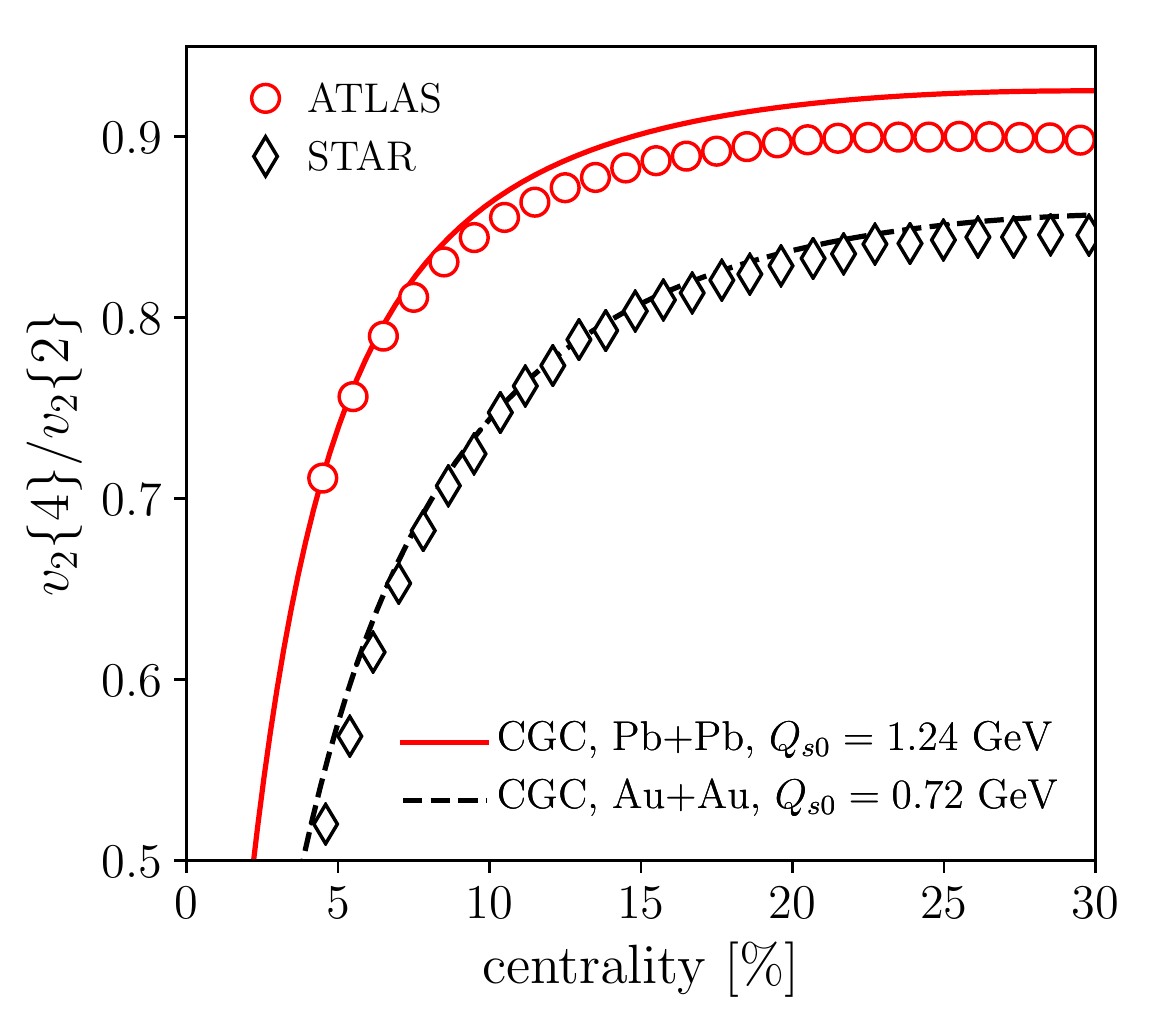}
    \caption{Ratio of the first two cumulants of elliptic flow, $v_2\{4\}/v_2\{2\}$. Symbols are STAR data (diamonds)~\cite{Adamczyk:2015obl} and ATLAS data (circles)~\cite{Aaboud:2019sma}. Lines are the results of the CGC model. Figure from~\cite{Giacalone:2019kgg}.}
    \label{fig:2}
\end{figure}

Moving on to the relative fluctuations of $v_2$, they also serve as a probe of the initial state since elliptic flow is a measure of the initial eccentricity, $\varepsilon_2$~\cite{Teaney:2010vd}, of the system, a quantity which originates from the fluctuating geometry of the initial energy-density field. In general, at a given centrality we have $v_2=\kappa\varepsilon_2$, where $\kappa$ is a response coefficient. The relative fluctuation of $v_2$ can be quantified by the ratio of the first two cumulant of its distribution, $v_2\{4\}/v_2\{2\}$, which is simply equal to $\varepsilon_2\{4\}/\varepsilon_2\{2\}$ because the coefficient $\kappa$ cancels in the ratio~\cite{Giacalone:2017uqx}. This allows us in particular to compare the relative $v_2$ fluctuations between RHIC and LHC without knowing how the coefficient $\kappa$ evolves with energy. In absence of fluctuations, the ratio is equal to unity, while $\varepsilon_2\{4\}/\varepsilon_2\{2\}<1$ for a fluctuating initial state. The deviation of this quantity from unity quantifies the amount of fluctuations in the system. I show in Fig.~\ref{fig:2} the ratio $v_2\{4\}/v_2\{2\}$ measured at both RHIC and LHC. STAR data are lower than ATLAS data. The initial state fluctuates more at RHIC.

Armed with this knowledge, we argue now that the energy evolution of the previous observables is captured by the model we introduced in Ref.~\cite{Giacalone:2019kgg}, which treats the initial condition of heavy-ion collisions as the energy density field produced immediately after ($\tau=0^+$) two sheets of Color Glass Condensate~\cite{Iancu:2003xm} cross each other. The statistics of energy-density fluctuations in this model was derived in Ref.~\cite{Albacete:2018bbv}. The local average (1-point function) of energy density reads:
\begin{equation}
\label{eq:1p}
\langle \rho({\bf s}) \rangle = \frac{4}{3g^2} Q_A^2({\bf s})Q_B^2({\bf s}),
\end{equation}
where ${\bf s}$ is a transverse coordinate, and $Q^2_{A/B}$ is the saturation scale (squared) of nucleus $A/B$, which we take proportional to the nuclear density integrated along the collision axis (\textit{thickness} function), usually denoted by $T_{A/B}$, with a coefficient in front, $Q^2_{s0}$, which gives the value of the saturation scale at the center of the nucleus. 
Fluctuations of energy density are, at leading logarithmic accuracy, given instead by the following short-range connected 2-point function:
\begin{equation}
\label{eq:2p}
    \nonumber \langle  \rho({\bf s}_1)\rho({\bf s}_2) \rangle - \langle \rho({\bf s}_1) \rangle \langle \rho({\bf s}_2) \rangle = \delta({\bf r})\xi({\bf s}),
\end{equation}    
with ${\bf s} = ({\bf s}_1+{\bf s}_2)/2$ and ${\bf r}={\bf s}_1-{\bf s}_2$, and
\begin{equation}
 \xi({\bf s}) = \frac{16\pi}{9g^4}  Q_{A}^2({\bf s}) Q_{B}^2({\bf s})
  \left[Q_{A}^2({\bf s})\ln\left(\frac{Q_{B}^2({\bf s})}{m^2}\right)
  +Q_{B}^2({\bf s})\ln\left(\frac{Q_{A}^2({\bf s})}{m^2}\right)\right],
\end{equation}
where $m$ is an infrared scale which cuts off the correlation of two color sources in the transverse plane. We shall use $m=0.14$~GeV, i.e., the pion mass. Let us study, then, the previous observables within this model.

The relative fluctuations of the total energy per unit longitudinal length, $E$, are equal to:
\begin{equation}
\label{eq:efluct}
    \frac{\sigma[E]}{\langle E \rangle} = \frac{\sqrt{\int_{\bf s} \xi({\bf s}) }}{\int_{\bf s} \langle \rho({\bf s}) \rangle } \propto \frac{1}{Q_{s0}},
\end{equation}
neglecting slowly-varying logarithms.
Now, fits of the Bjorken-$x$ evolution of structure functions measured at HERA indicate that the saturation scale evolves with collision energy as follows~\cite{Albacete:2014fwa}:
\begin{equation}
\label{eq:hera}
\frac{Q_s[{\rm LHC}]}{Q_s[{\rm RHIC}]} = \biggl(\frac{\sqrt{s_{\rm LHC}}}{\sqrt{s_{\rm RHIC}}}\biggr)^{0.14} \approx 1.5.
\end{equation}
This yields the same factor 1.5 obtained in the comparison between the relative fluctuations of multiplicity at RHIC and LHC in Eq.~(\ref{eq:mfluct}). Multiplicity fluctuations, hence, are quantitatively consistent with an $1/Q_s$ scaling, precisely as predicted by our model [Eq.~(\ref{eq:efluct})]. To visualize how the fluctuations of $E$ look like, we compute its distribution using the \texttt{magma} model~\cite{Gelis:2019vzt}, a Monte Carlo implementation of event-by-event profiles that fluctuate according to Eqs.~(\ref{eq:1p})~and~(\ref{eq:2p}). The distributions of $E$ for RHIC and LHC are shown in Fig.~\ref{fig:1}, on the right. Note that the different values of $Q_{s0}$ used in the calculations were fitted from anisotropic flow data~\cite{Giacalone:2019kgg}, and are consistent with the HERA scaling, Eq.~(\ref{eq:hera}).
Computing Eq.~(\ref{eq:efluct}), in central collisions we obtain:
\begin{equation}
    \frac{\sigma[E](b=0)}{\langle E \rangle (b=0)} \biggl|_{\rm RHIC} = 0.134, \hspace{50pt}  \frac{\sigma[E](b=0)}{\langle E \rangle (b=0)} \biggl|_{\rm LHC} = 0.088. 
\end{equation}
These numbers are essentially larger  by a factor 2 than those given in Eq.~(\ref{eq:mfluct}), as also suggested by the width of the tails of the histograms in Fig.~\ref{fig:1}. This is consistent with the fact that the distribution of $E$ will receive an important correction from the pre-equilibrium dynamics of the system during the first fm/$c$ of its evolution~\cite{Schlichting:2019abc}. The relative fluctuations of entropy at equilibrium are in fact expected to be significantly smaller than those of the initial energy~\cite{Giacalone:2019ldn}.

Finally, let us compute the fluctuations of elliptic flow. As anticipated, these originate from the fluctuations of the initial $\varepsilon_2$. Following Blaizot \textit{et al.}~\cite{Blaizot:2014nia}, and neglecting slowly-varying logarithms, the rms eccentricity due to fluctuations can be written as:
\begin{equation}
\label{eq:eccfluct}
     \frac{ \sqrt{ \int_{\bf s} |{\bf s}|^4 \xi({\bf s}) } }{ \int_{\bf s} |{\bf s}|^2 \langle \rho({\bf s}) \rangle } \propto \frac{1}{Q_{s0}}.
\end{equation}
The saturation scale appears in the denominator. Interestingly, this implies that $\varepsilon_2$ is larger at RHIC than at LHC. Nevertheless, the measured $v_2$ is smaller at RHIC, because the response $\kappa$ is strongly suppressed by the lower collision energy. Relative fluctuations, though, do not depend on $\kappa$ and can be genuinely compared. The CGC, thus, naturally predicts that they are larger at RHIC, in agreement with the experimental data shown in Fig.~\ref{fig:2}. Results on $v_2\{4\}/v_2\{2\}=\varepsilon_2\{4\}/\varepsilon_2\{2\}$ in the CGC are reported as lines in Fig.~\ref{fig:2}, and describe quantitatively the data.

In summary, the energy evolution of fluctuations provides a powerful probe of the initial state of nucleus-nucleus collisions. Experiments indicate that initial-state fluctuations in heavy-ion collisions are larger at RHIC than at LHC. This feature is both qualitatively and quantitatively captured by the CGC-inspired model of Ref.~\cite{Giacalone:2019kgg}, where fluctuations are inversely proportional to the saturation scale of the colliding nuclei.

G.G. M.L. and J.-Y.O. acknowledge funding from USP-COFECUB (grant Uc Ph 160-16, 2015/13). The work of F.G. and C.M. was supported in part by the Agence Nationale de la Recherche under the project ANR-16-CE31-0019-02. M.L.~acknowledges support from FAPESP projects 2016/24029-6, 2017/05685-2, 2018/24720-6, and project INCT-FNA Proc.~No.~464898/2014-5.


\begin{thebibliography}{99}

\bibitem{Adamczyk:2015obl} 
  L.~Adamczyk {\it et al.} [STAR Collaboration],
  Phys.\ Rev.\ Lett.\  {\bf 115}, no. 22, 222301 (2015)
  doi:10.1103/PhysRevLett.115.222301
  [arXiv:1505.07812 [nucl-ex]].

\bibitem{Aaboud:2019sma} 
  M.~Aaboud {\it et al.} [ATLAS Collaboration],
  [arXiv:1904.04808 [nucl-ex]].

\bibitem{Das:2017ned} 
  S.~J.~Das, G.~Giacalone, P.~A.~Monard and J.~Y.~Ollitrault,
  Phys.\ Rev.\ C {\bf 97}, no. 1, 014905 (2018)
  doi:10.1103/PhysRevC.97.014905
  [arXiv:1708.00081 [nucl-th]].
  
\bibitem{Teaney:2010vd} 
  D.~Teaney and L.~Yan,
  Phys.\ Rev.\ C {\bf 83}, 064904 (2011)
  doi:10.1103/PhysRevC.83.064904
  [arXiv:1010.1876 [nucl-th]].
  
\bibitem{Giacalone:2017uqx} 
  G.~Giacalone, J.~Noronha-Hostler and J.~Y.~Ollitrault,
  Phys.\ Rev.\ C {\bf 95}, no. 5, 054910 (2017)
  doi:10.1103/PhysRevC.95.054910
  [arXiv:1702.01730 [nucl-th]].
  
\bibitem{Giacalone:2019kgg} 
  G.~Giacalone, P.~Guerrero-Rodr\'iguez, M.~Luzum, C.~Marquet and J.~Y.~Ollitrault,
  Phys.\ Rev.\ C {\bf 100}, no. 2, 024905 (2019)
  doi:10.1103/PhysRevC.100.024905
  [arXiv:1902.07168 [nucl-th]].

\bibitem{Iancu:2003xm} 
  E.~Iancu and R.~Venugopalan,
  In *Hwa, R.C. (ed.) et al.: Quark gluon plasma* 249-3363
  doi:10.1142/9789812795533\_0005
  [hep-ph/0303204].

\bibitem{Albacete:2018bbv} 
  J.~L.~Albacete, P.~Guerrero-Rodr\'iguez and C.~Marquet,
  JHEP {\bf 1901}, 073 (2019)
  doi:10.1007/JHEP01(2019)073
  [arXiv:1808.00795 [hep-ph]].
  
\bibitem{Albacete:2014fwa} 
  J.~L.~Albacete and C.~Marquet,
  Prog.\ Part.\ Nucl.\ Phys.\  {\bf 76}, 1 (2014)
  doi:10.1016/j.ppnp.2014.01.004
  [arXiv:1401.4866 [hep-ph]].
  
\bibitem{Gelis:2019vzt} 
  F.~Gelis, G.~Giacalone, P.~Guerrero-Rodr\'iguez, C.~Marquet and J.~Y.~Ollitrault,
  arXiv:1907.10948 [nucl-th].
  
\bibitem{Schlichting:2019abc} 
  S.~Schlichting and D.~Teaney,
  arXiv:1908.02113 [nucl-th].
  
\bibitem{Giacalone:2019ldn} 
  G.~Giacalone, A.~Mazeliauskas and S.~Schlichting,
  arXiv:1908.02866 [hep-ph].

\bibitem{Blaizot:2014nia} 
  J.~P.~Blaizot, W.~Broniowski and J.~Y.~Ollitrault,
  Phys.\ Lett.\ B {\bf 738}, 166 (2014)
  doi:10.1016/j.physletb.2014.09.028
  [arXiv:1405.3572 [nucl-th]].

\end{thebibliography}
\end{document}